\documentclass[useAMS,usenatbib]{mn2e}
\usepackage{epsfig}

\newcommand{\prodd}{\displaystyle \prod}
\newcommand{\sumd}{\displaystyle \sum}
\newcommand{\ov}[1]{\overline{#1}}
\newcommand{\os}{_O}
\newcommand{\is}{_I}

\title[Practical planet prospecting]{Practical planet prospecting}

\author[S. Aigrain and M. Irwin]
  {S. Aigrain$^{1}$\thanks{E-mail: suz@ast.cam.ac.uk} and
   M. Irwin$^{1}$\thanks{E-mail: mike@ast.cam.ac.uk} \\
   $^{1}$Institute of Astronomy, University of Cambridge, Madingley
   Road, Cambridge, CB3 0HA, United Kingdom}

\begin{document}

\date{Accepted \ldots Received \ldots; in original form \ldots}

\pagerange{\pageref{firstpage}--\pageref{lastpage}} \pubyear{2002}

\maketitle

\label{firstpage}

\begin{abstract}
 
A number of space missions dedicated to the search for
exo-planets via the transit method, such as COROT, \emph{Eddington}
and \emph{Kepler}, are planned for launch over the next few
years. They will need to address problems associated with the
automated and efficient detection of planetary transits in light
curves affected by a variety of noise sources, including stellar
variability. To maximise the scientific return of these missions, it
is important to develop and test appropriate algorithms in advance of
their launch dates.

Starting from a general purpose maximum likelihood approach we discuss
the links between a variety of period and transit finding methods.
The natural endpoint of this hierarchy of methods is shown to be a
fast, robust and statistically efficient least-squares algorithm based
on box-shaped transits.

This approach is predicated on the assumption of periodic transits
hidden in random noise, usually assumed to be superposed on a flat
continuum with regular continuous sampling. We next show how to
generalise the transit finding method to the more realistic scenario
where complex stellar (micro) variability, irregular sampling and long
gaps in the data are all present.

Tests of this methodology on simulated \emph{Eddington} light curves,
including realistic stellar micro-variability, irregular sampling and
gaps in the data record, are used to quantify the performance.
Visually, these systematic effects can completely overwhelm the
underlying signal of interest.  However, in the case where transit
durations are short compared to the dominant timescales for stellar
variability and data record segments, it is possible to decouple the
transit signal from the remainder.

We conclude that even with realistic contamination from stellar
variability, irregular sampling, and gaps in the data record, it is
still possible to detect transiting planets with an efficiency close
to the idealised theoretical bound.  In particular, space missions
have the potential to approach the regime of detecting earth-like
planets around G2V-type stars.

\end{abstract}


\section{Introduction}
\label{sec:intro}

The discovery of the first exo-planet orbiting a Sun-like star was
announced almost a decade ago by \citet{mq95}. Since then
extraordinary progress has been made, and the number of planets
discovered to date is well beyond the hundred mark\footnote{see {\tt
www.exoplanet.org} \ or \ {\tt www.obspm.fr/encycl/encycl.html}.}. As
well as probing age-old questions such as the existence of life beyond
the Earth, these discoveries are fundamental to understanding how
planets and planetary systems form, and whether ours is a typical
one. The gaseous giant planets discovered so far have prompted a
re-thinking of planet formation theories due to their close-in and/or
eccentric orbits.

Among the various methods available to search for exo-planets, the
transit method presents a number of advantages. The most immediate are
that it allows direct determination of the planet's radius relative to
that of its parent star, the orbital inclination and, provided more
than one transit is observed, the orbital period. Combined with radial
velocity observations, a measurement of the planet mass free of the
$\sin i$ degeneracy can be obtained. The transit method also allows
the simultaneous monitoring of many thousands of target stars. This
multiplexing capability is a necessity, due to the stringent
requirement on the alignment of the orbit with the line of sight for
transits to occur. The first planet candidates tentatively detected
via the transit method have been announced over the last year or so
\citep{upz+02,uzs+02,msy+03,shl+03,dhk+03}, and one has received
tentative radial velocity confirmation \citep{ktj+03}.  The plethora
of ground-based searches currently underway (see \citealt{hor02} for a
review) is expected to yield hundreds of candidate transiting giant
exo-planets in the next few years.

However, terrestrial planets, capable of harbouring li\-quid water on
their surface, are beyond the reach of the methods used so
far. Detecting them is the goal of a number of planned
space-missions, such as the Franco-European satellite COROT
\citep{bag+03}, NASA's \emph{Kepler} \citep{bkl+03} and ESA's
\emph{Eddington}\footnote{COROT and \emph{Eddington} also include
asteroseismology programs.}  \citep{fav03}. These should push the
numbers of known exo-planets into the thousands.

The detection of a weak, short, periodic transit signal in noisy light
curves is a challenging task. The large number of light curves
collected make the automation and optimisation of the process a
necessity. This requirement is even stronger in the context of space
missions, which will collect even larger amounts of data and where
telemetry limitations will require as much of the processing to be
done on board as possible. A number of transit detection algorithms
have been implemented in the literature
\citep{ddk+00,ddb01,jcb02,upz+02,kzm02,af02,shl+03} and there has been
some effort to compare their respective performances in a controlled
fashion \citep{tin03}, but there is currently no widespread agreement
on the optimal method to use.

In a previous paper (\citealt{af02}, hereafter Paper I), a dedicated
Bayesian transit search algorithm was derived, based on the more
general period finding method of Gregory \& Loredo (\citealt{gl92b,
gre99}, hereafter GL92 and G99 respectively). Here we develop this
algorithm further and attempt to reconcile the apparent diversity of
the extant transit algorithms.  Starting from the original Gregory \&
Loredo prescription, which is based on a maximum likelihood (ML)
estimation for a periodic step-function model of unspecified shape,
appropriate sequential simplifications can be made.  We demonstrate
that the levels of the step-function bins -- which define the shape of
the detected event -- are not free parameters, their optimal values
being fully defined by the data. The use of Bayesian priors can be
dropped, given the lack of information currently available on the
appropriate form for these priors.  Finally, for detection purposes,
the model can be simplified to an unequal mark-space ratio square wave
with only one out-of-transit and one in-transit value. The algorithm
itself and its implementation are presented in Sect.~\ref{sec:deri}.
The performance has proved better than that of the previous version,
and the computational requirements have been significantly reduced.
Pursuing this simplification has also highlighted the similarities
between the previously published transit detection methods.

However, ML-based algorithms are only optimised for data containing
simple transits embedded in random noise (usually well approximated by
a Gaussian distribution).  Real transit search light curves will
contain instrinsic stellar variability of various amplitudes and
shapes. They will also suffer from irregular sampling, with frequent
large gaps in the coverage. Combined, these effects can pose a major
threat to our ability to detect planets. This problem is illustrated,
for the case of ground-based data, by recent data from the UNSW planet
search project using the Automated Patrol telescope at Siding Springs
observatory: in 5 nights of observations of the open cluster NGC6633,
nearly all of the 1000 brightest stars were found to be variable at
the millimag level \citep{apt03}. With the even higher precision
possible with upcoming space missions ($\sim0.1$~mmag), this problem
will become even more acute due to the sensitivity to additional
stellar activity-induced variability. Worries that this could
seriously impair the detection of terrestrial planets have led to the
development of variability filters \citep{jen02, caf03}, but these are
applicable only to data with regular sampling and no gaps. In
Sect.~\ref{sec:pre}, we introduce more generic filters applicable to
irregularly sampled data, or data with gaps (as expected for space
missions, due for example to telemetry drop-outs).  Performance
estimation results are discussed in Sect.~\ref{sec:perf}, and their
implications in Sect.~\ref{sec:discus}. Finally, Appendix~\ref{sec:ex}
contains details of how the simulated \emph{Eddington} light curves
used throughout the paper were generated.

\section[]{Maximum Likelihood-based algorithms}
\label{sec:deri}

\subsection{Maximum likelihood approach in the Gaussian noise case}
\label{sec:ml}

Transit searches are generally performed by comparing light curves to
a family of models with a common set of parameters, differing from
each other according to the different values used for these
parameters. The best set of parameters is identified by finding the
model most likely to have given rise to the observed data, i.e.\ the
model with the highest likelihood $L$.

If the noise in each data point $d_i$ is assumed to be Gaussian
(an assumption also valid for Poisson noise in the limit of large
numbers of photons), the likelihood can be written as the product of
independent Gaussian probability distribution functions:
\begin{equation}
  \label{eq:genL1}
  L = \prodd_{i=1}^{N} \left\{
        \frac{1}{\sqrt{2 \pi \sigma^2_i}} \ 
          \exp \left[ -\frac{(d_i - r_i)^2}{2 \sigma^2_i} \right] 
      \right\}
\end{equation}
where $d_i$ is the data value at time $t_i$ and $r_i$ is
the corresponding model value, $N$ is the total number of data points
and $\sigma_i$ the error associated with $d_i$. Eq.~(\ref{eq:genL1})
can be rewritten as:
\begin{equation}
  \label{eq:genL2}
  L = \left( \frac{1}{2 \pi} \right)^{(N/2)} \times \ 
        \prodd_{i=1}^{N} \left( \frac{1}{\sigma_i} \right) \times 
      \exp \left( -\frac{\chi^2}{2} \right)
\end{equation} 
where:
\begin{equation}
  \label{eq:genchi2}
  \chi^2 = \sumd_{i=1}^{N} \ \left[ 
             \frac{(d_i - r_i)^2}{\sigma^2_i} 
           \right]
\end{equation}
so that likelihood maximisation, in the case of Gaussian noise 
is equivalent to $\chi^2$ minimisation, since the
noise properties $\sigma_i$ are assumed to be known, i.e.\ fixed.

\subsection{The Gregory-Loredo method}
\label{sec:gl}

\begin{figure}
  \centering \epsfig{file=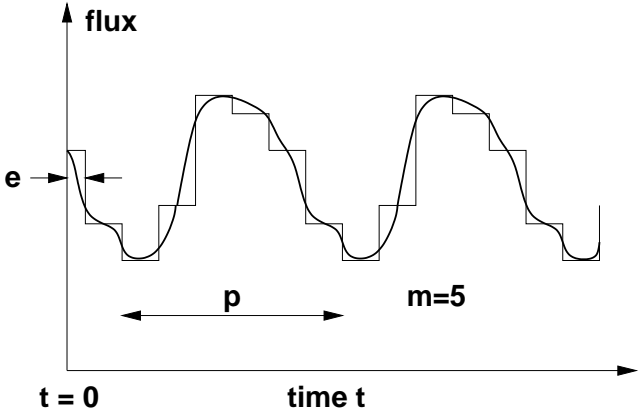,width=\linewidth}
  \caption{Schematic representation of the family of step-function 
    models used in the Gregory-Loredo method.}
  \label{fig:glmod}
\end{figure}

The generic method developed by Gregory \& Loredo (GL92, G99) to
detect periodic modulations in X-ray data, was used as the starting
point of the present work. This method is based on a Bayesian maximum
likelihood approach where the model consists of a periodic step
function with period $p$, and $m$ bins (labelled $1$ to $j$) of equal
duration $p/m$ (which can readily be generalised to unequal duration
bins if necessary). Each model is characterised by $p$, $m$, the epoch
$e$ (which is equal to the time $t$ at the start of the first bin) and
the individual bin levels $r_j$. Such a model is illustrated in
Fig.~\ref{fig:glmod}. The repartition of the data points into the $m$
bins is defined by:
\begin{equation}
  \label{eq:glbins}
  j_i = \mathrm{int} \left\{ 1 + m \left[\ 
          (t_i + p - e)~\mathrm{mod}~p \ \right] / p \right\}
\end{equation}
where $j_i$ is the number of the bin into which the $i^{\mathrm{th}}$
data point falls and $\mathrm{int}(x)$ is the largest integer
less than or equal to $x$.

For a given $m$, $p$ and $e$, the contributions from all possible
values for the individual bin levels $r_j$ are analytically integrated
over. Individual likelihoods are then computed at each point in the
$(m,p,e)$ parameter space. By marginalising over each parameter in
turn, one obtains a global posterior probability for the entire family
of periodic models. Marginalising over a parameter $\theta$ consists
of multiplying the (multi-dimensional) likelihood function by the
(assumed) prior probability distribution (Bayesian prior) for
$\theta$, then integrating over all values of $\theta$. This global
posterior probability can then be divided by the equivalent
probabilities for a constant and/or aperiodic model to give an odds
ratio, which is greater than $1$ if there is significant evidence for
periodicity. If this is the case, a posterior probability distribution
for each parameter $\theta$ can be computed by marginalising the
likelihood function over all the other parameters. The best value of
$\theta$ is that which gives rise to the maximum in the 1-D posterior
probability distribution for $\theta$. The interested reader is
referred to GL92 \& G99 for more details.

We discuss in the next section how this approach can be modified,
without loss of generality, to obviate the need for marginalising out
the $m$ variables $r_j$, corresponding to the values of each model
bin. This in turn leads to a very simple transit detection algorithm
for the special case of two discrete levels, of unequal duration,
applicable to most generic transit searches.

\subsection{Optimum $\chi^2$ calculation}
\label{sec:ochi2}

By directly maximising the likelihood, or in this case minimising
$\chi^2$, for any generalised step-function model, it is
straightforward to show that whatever the number and relative duration
of the bins, the optimal value for the bin levels $r_j$ can be
determined directly from the data given the other model parameters
$p$, $m$ and $e$. If we refer to the contribution from bin $j$ to the
overall $\chi^2$ as $\chi^2_j$, and define $J$ as the ensemble of
indices falling into bin $j$, we have:
\begin{equation}
  \label{eq:chi2j1}
  \chi^2_j = \sumd_{i \in J} \left[
               \frac{(d_i - r_j)^2}{\sigma^2_i} 
             \right]
\end{equation}
The value $\widetilde{r_j}$ of the model level $r_j$ that 
minimises $\chi^2_j$ is then simply given by the standard inverse
variance-weighted mean of the data inside bin $j$, since by setting
$\partial \chi^2_j / \partial r_j$ to zero we have:
\begin{equation}
  \label{eq:dc2}
  \frac{\partial \chi^2_j}{\partial r_j} = 
    2 \sumd_{i \in J} \left( \frac{r_j - d_i}{\sigma^2_i} \right) = 0
\end{equation}
hence:
\begin{equation}
  \label{eq:dmj}
  \widetilde{r_j} = \ov{d_j} 
                  = \left[ \sumd_{i \in J} \  \sigma^{-2}_i  \right]^{-1}
                      \sumd_{i \in J} \  d_i \sigma^{-2}_i 
\end{equation}
Substituting into Eq.~(\ref{eq:chi2j1}), $\chi^2_j$ now becomes:
\begin{equation}
  \label{eq:chi2j2}
  \widetilde{\chi^2_j} = \sumd_{i \in J} \left[ 
               \frac{ \left( d_i - \ov{d_j} \right)^2}{\sigma^2_i} \right]
\end{equation}
\noindent where $\widetilde{\chi^2_j}$ denotes the minimised value of
$\chi^2_j$. The contribution from each of the $m$ bins can be
simplified by expanding Eq.~(\ref{eq:chi2j2}):
\begin{equation}
  \label{eq:chi2j3}
  \widetilde{\chi^2_j} = \sumd_{i \in J} \left[ 
               \frac{d_i^2 - 2 d_i \ov{d_j} + {\ov{d_j}}^{~2}}
                    {\sigma^2_i}
             \right]
\end{equation}
\begin{equation}
  \label{eq:chi2j4}
  \widetilde{\chi^2_j} = \sumd_{i \in J} \ \frac{d_i^2}{\sigma_i^2} 
           - 2 \ov{d_j} \sumd_{i \in J} \ \frac{d_i}{\sigma_i^2} 
           + {\ov{d_j}}^{~2} \sumd_{i \in J} \ \frac{1}{\sigma_i^2} 
\end{equation}
From Eq.~(\ref{eq:dmj}) we have:
\begin{equation}
  \label{eq:dmj3}
  \sumd_{i \in J} \  \frac{d_i}{\sigma_i^2}  =
    \ov{d_j} \ \sumd_{i \in J} \  \frac{1}{\sigma_i^2} 
\end{equation}
so that:
\begin{equation}
  \label{eq:chi2j5}
  \widetilde{\chi^2_j} = \sumd_{i \in J} \  \frac{d_i^2}{\sigma_i^2} 
           - {\ov{d_j}}^{~2} \sumd_{i \in J} \ \frac{1}{\sigma_i^2} 
\end{equation}
The overall minimised $\chi^2$ over all bins is thus:
\begin{equation}
  \label{eq:chi2tot}
  \widetilde{\chi^2} = \sumd_{i=1}^{N} \  \frac{d_i^2}{\sigma^2_i}  
         - \sumd_{j=1}^{m} \left[ {\ov{d_j}}^{~2} 
             \sum_{i \in J} \  \frac{1}{\sigma^2_i}  \right]
\end{equation}

The first term in Eq.~(\ref{eq:chi2tot}) is entirely independent of the
model, and hence stays constant, so that only the second term needs to
be calculated for each set of trial parameters.
  
\subsection{Making use of the known characteristics of planetary transits}
\label{sec:af}

\begin{figure}
  \centering \epsfig{file=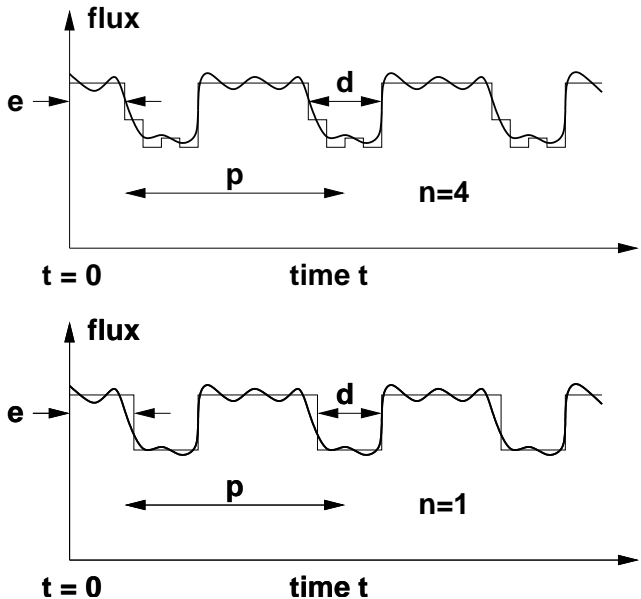,width=\linewidth}
  \caption{Schematic representation of the family of models used in Paper I 
    (top), and in the present paper (bottom).}
  \label{fig:newmod}
\end{figure}

The Gregory-Loredo method makes no assumptions about the shape of the
variations, and is fairly computationally intensive. However, when
trying to detect planetary transits, most of the information is
concentrated in a very small portion of the light curve. In a previous
paper \citep[][hereafter Paper~I]{af02}, we adapted the Gregory-Loredo
method to the planetary transit case by having one long out-of-transit
bin (bin $0$) and $n$ short in-transit bins (see
Fig.~\ref{fig:newmod}, top panel). The value of $n$ used was typically
$4$. For a given $n$, the parameters defining each candidate model are
then $p$, $e$, and the transit duration $d$. The likelihood
computation was carried out as described in G99.

This algorithm performed well when tested on simulated
data\footnote{The simulated light curves included some or no transits
and photon noise corresponding to the characteristics of the
\emph{Eddington} mission.}, but the likelihood calculation was still
computationally intensive. The odds ratio method was not used to
identify light curves showing significant evidence of transits, due to
considerations detailed in Paper~I. Instead, bootstrap simulations
containing hundreds of light curves with different realisations of the
same noise distribution, with and without transits, were used to
define optimised detection thresholds in terms of posterior
probability maxima.

A number of improvements have been made since the publication of
Paper~I:
\begin{enumerate}
\item Given the current state of exo-planet research, the use of
  Bayesian priors is not expected to contribute significantly to the
  performance of the algorithm. The information available on period
  and duration distributions is relatively scarce for giant planets,
  and non-existent for terrestrial planets. The priors used in Paper I
  were generic and mostly identical to those used by G99 for X-ray
  pulsars, rather than specifically optimised for transit searches.
\item Using the $\chi^2$ rather than the likelihood as a detection
  statistic, and implementing the calculation as outlined in
  Sect.~\ref{sec:ochi2}, significantly reduces the computational
  requirements of the detection process.
\item The shape of most planetary transits is sufficiently simple
  that, for detection purposes (as opposed to detailed parameter
  estimation), a single in-transit bin, as illustrated in
  Fig.~\ref{fig:newmod} (bottom panel) provides enough information.  A
  significant advantage of this simplification is that it makes the
  method far more robust and capable of coping with real data, and all
  its concomitant problems, with negligible loss in detection
  efficiency.
\item Once a detection is made, a shape-estimation phase with either a
  large value of $n$, or by detailed model fitting of the phase folded
  light curve, can be implemented. As the dependency of transit shapes
  as a function of the stellar and planetary parameters is relatively
  well-known, Bayesian priors may have a part to play in this phase.
  This is, however, outside the scope of the present paper.
\end{enumerate}

\subsection{$\chi^2$-minimisation with a box shaped transit.}
\label{sec:chi2box}

The algorithm used in the present paper evolved from that of Paper~I,
taking into consideration the points listed in Sect.~\ref{sec:af}.
The model therefore consists of one out-of-transit bin and a single
level in-transit bin.  (Although this simplification may seem
disingenuous, by suitably pre-processing, or adaptively filtering, the
signal to remove intrinsic stellar variability, this is a valid
approximation to transit detection in practice.)  All the data points
falling into the out-of-transit bin form the ensemble $O$, while those
falling into the in-transit bin form the ensemble $I$. No Bayesian
priors are used. Adapting Eq.~(\ref{eq:chi2tot}) to this model gives:
\begin{equation}
  \label{eq:chi2n}
  \widetilde{\chi^2} = \sumd_{i=1}^{N} \  \frac{d_i^2}{\sigma^2_i} 
             - {\ov{d\os}}^{~2} \sumd_{i \in O} \ 
                                  \frac{1}{\sigma^2_i} 
             - {\ov{d\is}}^{~2} \sumd_{i \in I} \ 
                                  \frac{1}{\sigma^2_i} 
\end{equation}
Provided the transits are shallow and of short duration (i.e.\ the most
common case), the ensemble $O$ contains the vast majority of the data
points, so that $\ov{d\os} \approx \ov{d}$ (where $\ov{d}$ is the
weighted mean of the entire light curve). Substituting this
approximation into Eq.~(\ref{eq:chi2n}):
\begin{equation}
  \label{eq:chi2sp}
  \widetilde{\chi^2} \approx \sumd_{i=1}^{N} \left\{ 
                   \frac{d_i^2}{\sigma^2_i} 
                   -  \frac{\ov{d}^2}{\sigma^2_i}
                 \right\}
               - {\ov{d\is}}^{~2} \sumd_{i \in I} \ \frac{1}{\sigma^2_i} 
\end{equation}
The first two terms in Eq.~(\ref{eq:chi2sp}) are constant. The
minimisation of $\chi^2$ is therefore achieved by maximising the
detection statistic $Q$, given by:
\begin{equation}
  \label{eq:Q1}
  Q = {\ov{d\is}}^{~2} \sumd_{i \in I} \  \frac{1}{\sigma^2_i}
\end{equation}
which can also be expanded as:
\begin{equation}
  \label{eq:Q2}
  Q = \left ( \sumd_{i \in I} \  \frac{d_i}{\sigma^2_i} \right)^2 
      \left ( \sumd_{i \in I} \  \frac{1}{\sigma^2_i} \right ) ^{-1}
\end{equation}
If the light curve is robustly ``mean-corrected'' prior to running the
algorithm, such that $d_i$ is replaced by $\Delta d_i$, $\ov{d\is}$
becomes $\ov{\Delta d\is}$, the depth of the model transit. This
results in a further simplification where the only free parameters are
now the phase, period, and duration of the transit, since the depth is
determined given the other three.  It is also apparent that $Q$ is
simply equal to the square of the in-transit signal-to-noise ratio.
This is easier to see in the case where $\sigma_i = \sigma$ for all
$i$ (a good approximation to the case for space
data). Eq.~(\ref{eq:Q2}) then becomes:
\begin{equation}
  \label{eq:Q3}
  Q = \left( \sumd_{i \in I} \Delta d_i \right)^2 
      \left( n_I \sigma^2 \right)^{-1}
    = \left( \sumd_{i \in I} \frac{\Delta d_i}{n_I} \right)^2
      \frac{n_I}{\sigma^2}
\end{equation}
where $n_I$ is the number of points in $I$, and $\sum_{i \in I} \Delta
d_i / n_I$ is the mean of the in-transit points, i.e.\ the model
transit depth (the weighting being unnecessary in that case).
 
Eq.~(\ref{eq:Q3}) is used when the errors are constant, or when no
individual error estimates are available for each data point. In the
latter case, the Median Absolute Deviation (MAD) of the dataset is
used to estimate $\sigma$, as this is more robust to outliers than a
simple standard error estimate \citep{hmt83}.
For a Gaussian distribution $\sigma_{rms} = 1.48 \times \mathrm{MAD}$
and this factor is used throughout to scale the MAD sigmas.  If
individual error estimates are available, Eq.~(\ref{eq:Q2}) provides a
more precise estimate of $Q$ at the cost of a slight increase in
computation time.

If the noise is Gaussian, a theoretical signal-to-noise threshold
(i.e.\ $Q$ threshold) can in principle be computed a priori to
keep the false alarm rate below a certain value \citep{jcb02}.

\subsection{Comparison with other transit search techniques}
\label{sec:compar}

In following through the steps of the previous sections our prime
motives were to modify a general purpose Bayesian periodicity
estimation algorithm to make it simpler, faster and more robust. In so
doing we have arrived at a very similar formulation to that developed
by other authors, though the details of the implementation differ.
For example, \citet{kzm02} derived and tested a box-fitting method
(BLS) similar to the present algorithm on simulated ground based data
with white noise, and showed that significant detections followed for
in-transit signal-to-noise ratios greater than $6$.

\citet{shl+03} used a transit finding algorithm based on a matched
filter technique.  After identifying and removing large amplitude
variable stars they generated model light curves consisting of a constant
out-of-transit level and a single in-transit section.  The models were
generated for a series of transit durations and phases, and a
$\chi^2$-like measure was then used to select the best model (indeed
their Eq.~3 is essentially a special case of the method derived in
Sect.~\ref{sec:ochi2} for single transits).

\citet{uzs+02}, who have claimed the first direct detections of
transiting planetary candidates, also implemented a version of
the BLS algorithm and noted that it was much more efficient than their
own algorithm based on ``a simple cross-correlation with an error-less
transit light curve'' \citep{upz+02}.

In a comparison of several transit finding algorithms, \citet{tin03}
found that matched filters and cross-correlation gave the best results
compared with progressively more general methods ranging from BLS,
through Deeg's method \citep{ddk+00} to Defa\"y's \citep{ddb01}
Bayesian approach.The fact that matched filters and cross-correlation
methods give good results is hardly surprising, and can easily be
deduced from the $\chi^2$ minimisation developed in
Sect.~\ref{sec:ochi2}.  Examination of Eq~(\ref{eq:chi2j1}) shows that
the dominant term is the cross-term $\sum d_i r_j / \sigma_i^2$, which
needs to be maximised. The first term is a constant for a given
dataset, while the final model term should have much smaller
influence. The cross-term is exactly a generalised cross-correlation
function and also identical to a matched filter. The more general
methods suffer from the added complexity of the underlying model,
which through the Bayesian view of Occam's Razor, reduces the
tightness of the posterior probability distribution of the parameter
estimation. What is however surprising, is that the BLS method did not
give at least as good a result as the matched filter and
cross-correlation methods.  We would expect the BLS method to have
similar performance to the matched filter as it is mathematically
almost identical.

\subsection{Optimised parameter space coverage}
\label{sec:sp}

The formulation of the detection statistic presented in
Sect.~\ref{sec:chi2box} is fully defined given only the dataset and
the start and end times of each model transit. The model parameters
are thus the duration $d$, period $p$ and epoch/phase $e$ (defined for
our purposes as the time at the start of the first transit in the
dataset).

The range of expected transit durations is relatively small -- from a
few hours for close-in, rapidly orbiting planets, to almost a day for
the most distant planets transiting more than once within the
timescale of the planned observations. A simple discrete sampling
prescription can therefore be adopted for the duration without leading
to large numbers of trial values.  One option is to choose the step
$\delta d$ between successive trial durations to be approximately
equal to the average time step $\delta t$ between consecutive data
points. This ensures that models with the same period and epoch and
neighbouring trial durations differ on average by $\sim 1$ data point
per transit.  However, if the observation sampling rate is high -- a
sampling rate of 10~min is envisaged for most targets for
\emph{Eddington} in planet-finding mode \citep{fav03} -- a larger step
in duration can be used, provided it is smaller than the shortest
significant feature in the transit, namely the ingress and egress,
which have typical durations of $\sim 30$~min.

The period sampling prescription is designed to ensure that the error
in the phase (or equivalently epoch) of the last model transit in the
light curve is smaller than a prescribed value. Capping the error on the
period (by using a constant trial period step) is not sufficient, as
the error on the epoch of the $n^{\mathrm{th}}$ transit will be $n$
times the error on the epoch of the first. This would lead to a larger
overall error for shorter periods, where the number of transits in the
light curve is large, thus introducing a bias in the distribution of
detection statistic with period. This bias is not present if one uses
a constant step in trial frequency. Defining the relative frequency
$\nu = T/p$, $T$ being the total light curve duration, the phase of an
event occurring at time $t$ is given by $\theta = 2 \pi t / p = 2 \pi t
\nu / T$, so that for the last transit in the light curve $\theta
\approx \theta_{\mathrm{max}} = 2 \pi \nu$.  A fixed step in $\nu$
thus leads to a fixed error in $\theta_{\mathrm{max}}$. By trial and
error, a value of 0.05 was found to be suitable for $\delta \nu$.
  
One caveat in the case of space missions with high sampling rates
lasting several years, is that the above prescription can lead to very
large numbers of trial periods. This implies that the overall
algorithm must be extremely efficient. Some steps taken to optimise
the efficiency are described below. 

The phase, or epoch step interval, is set to the average sampling rate
of the data since by so doing one can generate the phase information
at no extra computational cost using an efficient search algorithm,
detailed below.

\subsection{A weighting scheme to account for non-continuous sampling}
\label{sec:weight}

A further complication stemming from irregular sampling and from the finite
duration of each sample, is that data points nominally corresponding to a 
time outside a transit may correspond partly to the out-of transit bin and 
partly to the in-transit bin. To account for this, the indices of points 
falling either side of the transit boundaries are also stored and included is
the calculation of $Q$, but with a weight which is $< 1$ and is
inversely proportional to the interval between the time corresponding
to the data point and the start/end time of the transit. This
weighting scheme is particularly important for data with irregular
sampling where transits might fall, for example, at the end of a night 
of ground-based observations, or even with spaced-based observations
during a gap in the temporal coverage.

\subsection{Speeding up the algorithm}
\label{sec:speed}

By far the most time consuming operation in computing $Q$ and finding
the set of parameters which maximises it, is the identification of the
in-transit points, which must be identified for each model $d$, $p$ and $e$.
If one is dealing with a large number of light curves sharing the same
observation times, it is more efficient to process many light curves
simultaneously and compute $Q(d,p,e)$ for the entire block of
light curves for each set of parameters, as follows. For each trial
period, the time array is phase-folded. At a given trial duration, the
in-transit points are identified for the first trial epoch, by
stepping through the folded time array one element at a time until the
start time of the transit is reached, and then continuing, storing the
corresponding indices, until the end time of the transit is reached.
$Q(d,p,e)$ is then computed and stored for each light curve. When
moving to the next trial epoch, one steps backward through the folded
time array from the end time of the old transit (which is stored
between successive trial epochs) until the start time of the new
transit is found. One then steps forward through the time array,
storing the indices, until the end time of the new transit is reached.
$Q(d,p,e)$ is then computed and stored, and the epoch incremented, and
so forth.

This minimises the overall number of calculations needed. As the number of
in-transit points is the same for all light curves and $\sigma$ only
needs to be computed once per light curve (in the constant error
case), this leaves only the sum of the in-transit points to be
computed once per set of parameters and per light curve.  The optimum
number of light curves to process simultaneously depends on the amount 
of memory available.

A further speed increase is obtained by noting the redundancy within
the computation of $Q$ for a range of phase/epoch and period trial
values.  Breaking down the search to a two-stage process consisting of
a single transient event detector (essentially a matched filter stage)
followed by a multiplexed period/phase search, removes the inner loop
summation of data from the main search and gives a factor of $\sim 10$
improvement in execution time.

Example run-times computed using a laptop equipped with a 1.2~GHz
Pentium~IV processor with 512~MB of RAM are as follows.  The light
curves consisted of 157\,680~floating point numbers, i.e.\ each
was $\sim 630$~KB in size. The trial period and duration ranges were
180 to 400~d and 0.5 to 0.7~d respectively. These ranges are roughly
appropriate to search for transits of planets in the habitable zone of
a Sun-like star, and correspond to a total number of tested $(p,d,e)$
combinations of $\sim 5\times 10^7$.  After finding the optimal number
of light curves to search simultaneously, the runtime per light curve
was $\sim 4$~s.

Note that close-in planets with periods below the range included in
this simulation are, of course, of interest, so that lower trial
periods (and hence lower trial durations) would also be included when
searching for transits in real data, thereby increasing the
runtime. As the trial period range is increased, the number of trial
periods becomes prohibitively large due to the use of even sampling in
frequency space (see Sect.~\ref{sec:sp}): this leads to very small
trial period steps at the low period end of the range if the steps are
to be kept reasonable at the high period end of the range. This can be
remedied by splitting the required range of trial periods and running
the algorithm separately for each period interval. The runtime
increases linearly with the number of trial durations.

\section{Pre-processing irregularly sampled data}
\label{sec:pre}

\begin{figure}
  \centering
  \epsfig{file=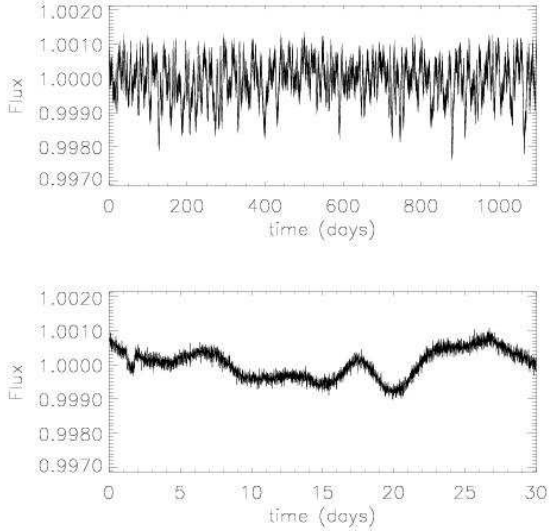, width = 0.9\linewidth}
  \caption{Simulated \emph{Eddington} light curve for a $V=13$
    solar-age G2V star orbited by a $2~R_{\oplus}$ planet with a
    period of 1~yr (see Appendix~\ref{sec:ex} for details). Top panel:
    Entire light curve. Bottom panel: First 30~days, with a transit
    1.5~d after the start. The flux values shown have been normalised
    to have a mean of 1.}
  \label{fig:lc}
\end{figure}

Intrinsic variability from the planet host star is expected to be the
dominant noise source for space-based planetary transit searches, and
for ground-based searches in the case of active stars. As an example,
we use throughout the present section a light curve simulated
according to the planned characteristics of the \emph{Eddington}
mission, containing stellar variability, planetary transits and photon
noise. The procedure used to generate this light curve is described in
more detail in Appendix~\ref{sec:ex}. The light curve, shown in
Fig.~\ref{fig:lc}, corresponds to a solar-age G2V star with apparent
magnitude $V=13$, containing transits of a $2~R_{\oplus}$ planet which
last $\sim 13$~hr and have a period of 1~yr. It has a sampling of
10~min and a duration of 3~yr.

Intrinsic stellar variability can seriously impede the detection of
terrestrial planets by missions such as \emph{Eddington} and
\emph{Kepler}.  However, it is possible to disentangle the planetary
transit signal from other types of temporal variability if the two
have sufficiently different temporal characteristics.  To illustrate
this we show the power spectra of the different components
contributing to the light curve mentioned above in
Fig.~\ref{fig:psnspec}.  Although the power contained in the transit
signal is small compared to both stellar and photon noise components
(and would be even smaller for the case of an Earth-size planet), it
retains significant power for frequencies higher than $\sim 1~\mu$Hz,
where the stellar signal starts to drop off steeply.  As long as this
condition is fulfilled (i.e.\ if the stellar variability occurs on
sufficiently long timescales), one should be able to separate and
detect the transits.  Furthermore, in the case of multiple transits,
the regular period of the transits also helps constrain the Fourier
space occupancy of the transit signal with respect to the stellar
signal.

\subsection{Wiener or matched filtering approach}
\label{sec:oldf}

\citet{caf03} demonstrated how use of an optimal filter can
simultaneously pre-whiten and enhance the visibility of transits in
data dominated by stellar variability.  The Fourier-based method
presented there is also closely related to a minimum mean square error
(MMSE) Weiner filter.  However, even for space-based missions
uneven sampling of the data will occur. In these real-life cases,
irregularly sampled data implies that standard Fourier methods are no
longer directly applicable and a more general technique is required.

\begin{figure}
  \centering
  \epsfig{file=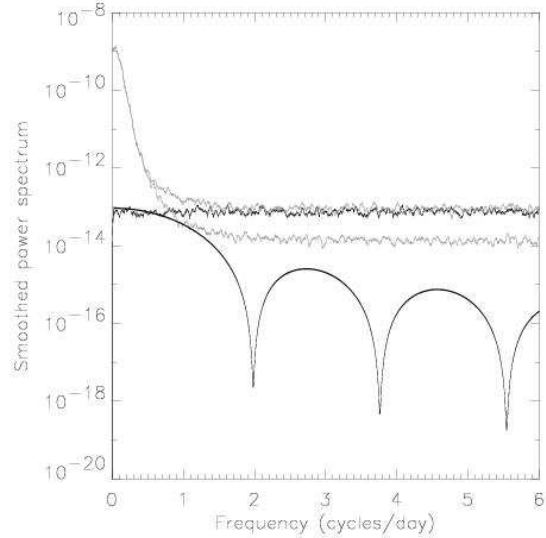, width = 0.9\linewidth}
  \caption{Power spectrum of the light curve shown in
    Fig.~\ref{fig:lc} (upper grey line). Lower grey line: stellar
    variability only. Lower black line: transits only (3
    transits). Upper black line: photon noise. The power spectrum is
    dominated by stellar variability at low frequencies and by photon
    noise at high frequencies.}
  \label{fig:psnspec}
\end{figure}

To gain some insight to the problem consider the general case of
intrinsic stellar variability, with the received signal $x(t)$ is
composed of the three components:
\begin{equation}
  \label{eq:3comp}
  x(t) = s(t) + r(t) + n(t)
\end{equation}
\noindent where $s(t)$ is the intrinsic time variable stellar light
curve, $r(t)$ is the transiting planet signal, and $n(t)$ denotes the
measurement plus photon noise, which we can take to be random (and
Gaussian in the cases of interest here)\footnote{Strictly speaking,
the $1^{\rm st}$ two terms in Eq.~(\ref{eq:3comp}) should be
multiplicative, but in the limit of low amplitude variability and
shallow transits, an additive combination is a very good
approximation.}.  Each component is statistically independent, hence
the expected power spectrum $\Phi(\omega)$ of the received signal is
simply given by:
\begin{equation}
  \label{eq:phi}
  \Phi(\omega) = \left<|S(\omega)|^2\right> + \left<|R(\omega)|^2\right> + 
                 \left<|N(\omega)|^2\right> 
\end{equation}
and in the case of random, or white, noise
$\left<|N(\omega)|^2\right>$ is a constant, hence guaranteeing
positivity of the right hand term.  This also highlights in a natural
way a justification for the somewhat arbitrary constant in Eq.~(6) in
\citet{caf03} and how its value is related to the expected noise
properties (although it would be more natural to implement it as a
lower bound).  However, as outlined below there is a simpler way to
implement their technique without the need for the additional
constant.
\begin{figure}
  \centering
  \epsfig{file=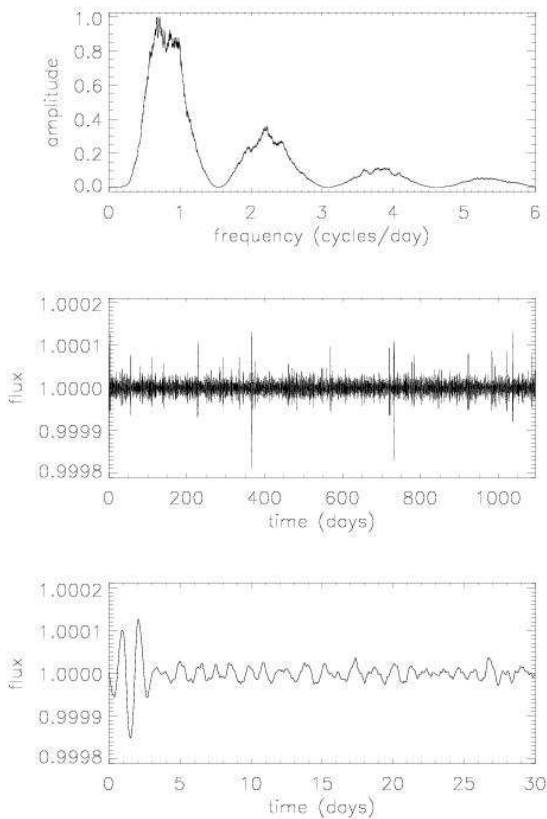, width = 0.9\linewidth}
  \caption{Top panel: Wiener filter constructed using the light curve
    shown in Fig.~\ref{fig:lc} and a reference box-shaped transit of
    duration 0.65~d . Middle panel: Filtered light curve. Bottom
    panel: Idem, $1^{\rm st}$ 30 days, with a transit 1.5~d after the
    start.}
  \label{fig:wiener}
\end{figure}

A standard MMSE Wiener filter attempts to maximise the signal-to-noise
in the component of interest, in this case $r(t)$, by convolving the
data with a filter, $h(t)$, constructed from the ratio of the
cross-spectral energy densities between observation and target, such
that:
\begin{equation}
  \label{eq:filt}
  x'(t) = h(t) \otimes x(t) \hspace{15mm}  X'(\omega) = H(\omega) \ X(\omega) 
\end{equation}
and (using $^*$ to denote complex conjugate):
\begin{equation}
  \label{eq:wien}
  H(\omega) = \frac{\left<R(\omega) X(\omega)^*\right>}
                   {\left<X(\omega) X(\omega)^*\right>}
            = \frac{\left<|R(\omega)|^2\right>}
                   {\left<|X(\omega)|^2\right>}
\end{equation}
for a long enough run (a fair sample) of observations.  In practice
the only example we have of $x(t)$ is often singular, implying that
the best estimate of the denominator is simply the observed power
spectrum $\Phi(\omega)$, subject to the constraint of positivity
imposed by the implicit $\left<|N(\omega)|^2\right>$ term. Such a
filter is illustrated in Fig.~\ref{fig:wiener}: the top panel shows the
filter, constructed using the Fourier transform of the light curve
shown in Fig.~\ref{fig:lc} and a box-shaped reference transit of
duration $0.65$~d, and the bottom two panels show the filtered light
curve.

\begin{figure}
  \centering
  \epsfig{file=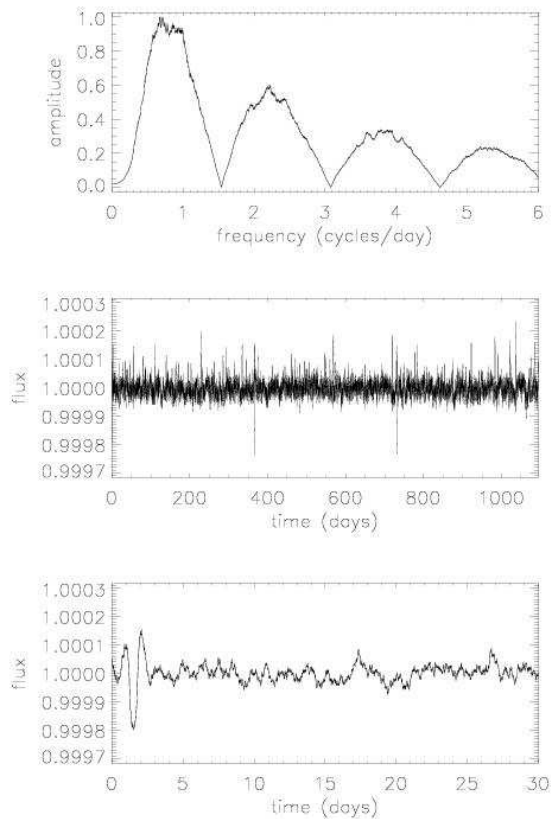, width = 0.9\linewidth}
  \caption{Top panel: Matched filter constructed using the light curve
    shown in Fig.~\ref{fig:lc} and a reference box-shaped transit of
    duration 0.65~d. Middle panel: Filtered light curve. Bottom
    panel: Idem, $1^{\rm st}$ 30 days, with a transit 1.5~d after the
    start.}
  \label{fig:wsqrt}
\end{figure}

This should be contrasted with the pre-whitened matched detection
filter employed by \citet{caf03}, illustrated in Fig.~\ref{fig:wsqrt}
(using the same layout as Fig.~\ref{fig:wiener}), and which can be
written in the form:
\begin{equation}
  \label{eq:pap2}
  X'(\omega) = H(\omega) \ X(\omega)              = \frac{X(\omega)}
                 {\left<|X(\omega)|\right>} \ \left<|R(\omega)|\right>
\end{equation}
and hence is equivalent to reconstructing the data using just the
phase of the input signal Fourier transform modulated by the amplitude
spectrum from the expected transit shape (see
Fig.~\ref{fig:wsqrt_phase}). Viewing the problem in this way removes
the need for the additional constant in their Eq.~(6) and emphasises
the two stage nature of the filtering. The pre-whitening suppresses
the stellar variability component, while the matched filter is
directly equivalent to the $n=1$ ML case presented in
Sect.~{\ref{sec:deri}.

\begin{figure}
  \centering 
  \epsfig{file=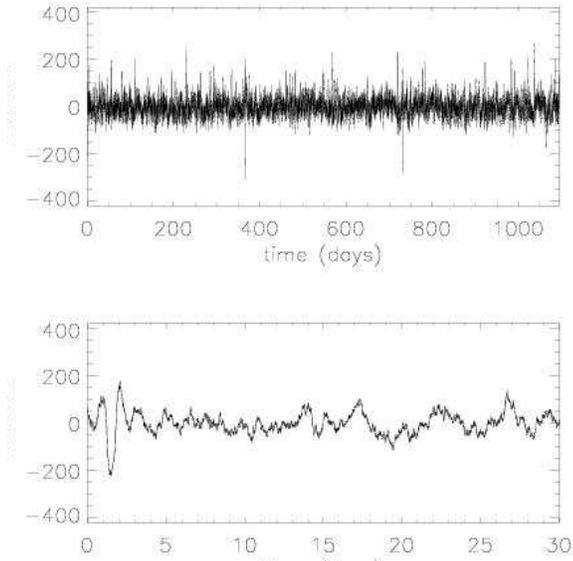, width = 0.9\linewidth, 
          bbllx = 57, bblly = 51, bburx = 540, bbury = 527, clip =}
  \caption{As Fig.~\ref{fig:wsqrt}, but the filtered light curve was
    obtained by modulating the phase of the Fourier transform of the
    data by the amplitude spectrum of the reference transit
    signal. The filter was omitted as it is effectively identical to
    that shown in Fig.~\ref{fig:wsqrt}. Comparing, visually, the
    amplitude, shape and timescale of the variations in the filtered
    data with the bottom two panels of Fig.~\ref{fig:wsqrt} confirms
    that this gives very similar results to the matched filter
    approach.}
  \label{fig:wsqrt_phase}
\end{figure}

In practice, transit searching can be based directly on the output of
the filtering, or preprocessing can be used to decouple the stellar
variation estimation from the transit search phase, which then
proceeds using the methods outlined in Sect.~\ref{sec:deri}, since the
problem has been reduced to the simpler one of transit detection in
random noise. (In either case, detailed investigation of the transit
depth and shape involves phase folding, unfiltered data, and local
modelling.)

Either of these preprocessing filters works well in the case of
regularly sampled data with no gaps and with a reasonable separation
between the signatures of the Fourier components of the transits and
the stellar variability. In Figs.~\ref{fig:wiener}, \ref{fig:wsqrt} \&
\ref{fig:wsqrt_phase}, the transits are distinctly visible in the
filtered light curve. The results in terms of transit detection
performance using either method are very similar. For simplicity, the
matched filter approach, rather than the Wiener filter, is used in the
remainder of this paper.

\begin{figure}
  \centering
  \epsfig{file=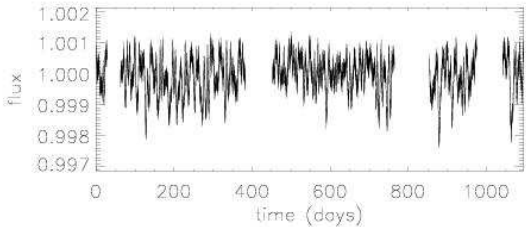, width = 0.9\linewidth}
  \caption{Simulated light curve with data gaps. Four arbitrarily
    chosen sections were removed from the light curve shown in
    Fig.~\ref{fig:lc}. Note that for this test the gaps were chosen to
    avoid the transit regions for comparison purposes.}
  \label{fig:lc_g}
\end{figure}

\begin{figure}
  \centering
  \epsfig{file=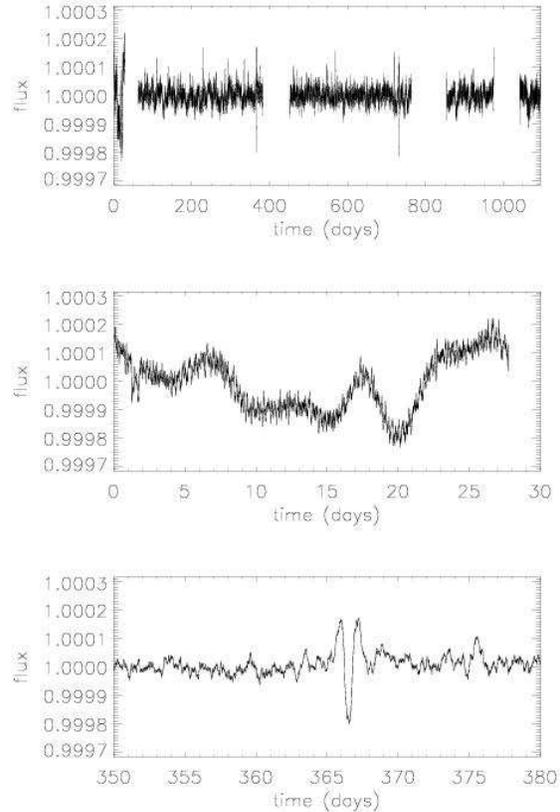, width = 0.9\linewidth}
  \caption{Results of applying the matched filter independently to the
    5 unbroken intervals of the light curve shown in
    Fig.~\ref{fig:lc_g}. Top panel: entire filtered light
    curve. Middle panel: $1^{\rm st}$ 30~days. Bottom panel: another
    30~d section centred on the second transit (at 366.5~d). See text
    for an explanation.}
  \label{fig:wsqrt_gi}
\end{figure}

However, real data, even space-based, suffers from irregular sampling
and the presence of significant gaps. Fourier domain methods cannot be
directly applied to irregularly sampled data, but it is possible to
treat regularly sampled data with gaps as a series of $n$ independent
time series, and to filter them separately. To test this, four
arbitrarily chosen sections were removed from the light curve shown in
Fig.~\ref{fig:lc} (see Fig.~\ref{fig:lc_g}). The matched filter was
then applied to the five unbroken intervals separately, and the
results are shown in Fig.~\ref{fig:wsqrt_gi}. Though the filtering is
effective on relatively long sections of data (bottom panel) it is not
successful for short intervals (middle panel), even if they are
significantly longer than the transit duration. This is because the
power spectrum of the stellar noise is estimated from the data in
order to construct the filter. For this to be successful, the data
segment needs to be at least twice as long as the longest significant
timescale in the star's variability, which is either the rotation
period or the long end of the starspot lifetime distribution
\citep{afg04}. In the case of the G2V star used in the simulations,
the minimum data segment length for which the filtering was successful
was $\sim 60$~days (last data segment in Fig.~\ref{fig:wsqrt_gi}),
consistent with a rotation period of $\sim 30$~days for such a star.

It is therefore necessary to find other means of coping with this
additional complexity. We have investigated two alternative
approaches: one based on a least-squares generalisation of the Fourier
filtering approach; the other based on a general purpose iterative
clipped non-linear filter.  In both cases we use the preprocessing to
attempt to remove the stellar signature, as much as possible, prior to
invoking the transit detection methods developed in
Sect.~\ref{sec:deri}.

\subsection{Least-squares filtering}
\label{sec:lsf}

For a long run of regularly sampled data, a discrete Fourier transform
asymptotically approaches a least-squares fit of individual sine and
cosine components (see e.g.\ \citealt{bre88}). This naturally suggests
an extension of the approach described in Sect.~\ref{sec:oldf} to the
case of irregularly sampled data. An analogous situation occurs in the
generalisation of the periodogram method to Fourier estimation of
periodicity; using generic least-squares sine curve fitting is a more
flexible alternative \citep{bw71}. This allows the case of gaps in the
data, or more generally irregular sampling, to be dealt with in a
consistent and simple manner.

\begin{figure}
  \centering
  \epsfig{file=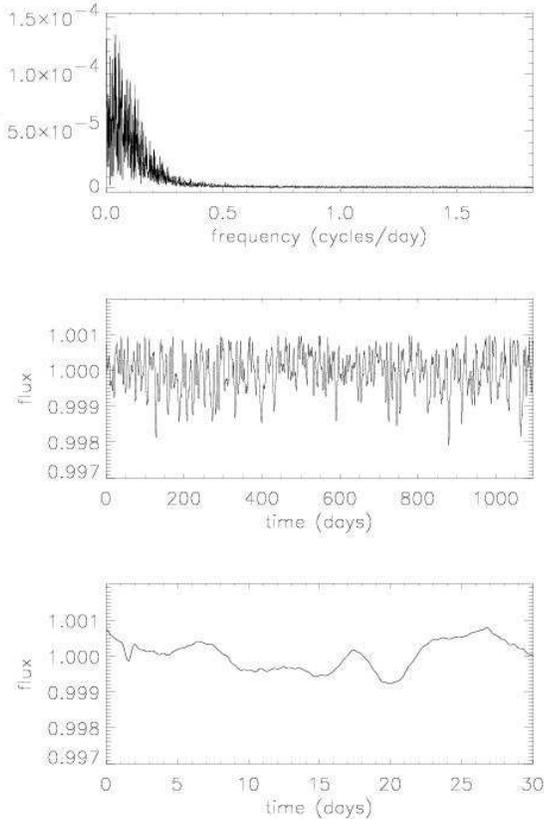, width = 0.9\linewidth}
  \caption{Top panel: ``Power spectrum'' (i.e.\ coefficients $a_k$
    versus frequency) obtained by the least-squares fitting method for
    the light curve shown in Fig.~\ref{fig:lc}. Middle panel:
    Reconstructed light curve, obtained by summing over the fitted
    sine-curves up to a frequency of $\sim 1.8$~cycles/day. Bottom
    panel: $1^{\rm st}$ 30~days of the reconstructed light curve.}
  \label{fig:lsfit}
\end{figure}

The procedure is basically identical to that employed for the Wiener
filter described in the previous section, but the calculation of the
Fourier transform, or power spectrum, of the received signal is
replaced by an orthogonal decomposition of this signal into sine
components whose amplitude, phase and zero-point are fitted by
least-squares.  Each of the components has the form:
\begin{equation}
  \label{eq:comp}
  \psi_k(t) = a_k \sin \left( 2 \pi k t / T + \phi_k \right)
\end{equation}
where $T$ is the time range spanned by the data.  The number of
components to fit can be chosen such than the maximum frequency fitted
is equal to some fraction of the Nyquist frequency, but for this one
must define an equivalent sampling time $\delta t$. In the case of
regular sampling with gaps, $\delta t$ is simply the time sampling
outside the gaps. In the case of irregularly sampled data the
definition of $\delta t$ is more open ended.  However, provided that
the sampling is close to regular, a good approximation will be the
average time step between consecutive data points -- keeping in mind
that any significant gaps should be excluded from the calculation of
this average.  The potentially highest frequency component should then
have frequency $\approx 1/ \left( 2 \delta t \right)$, although in
practice a much lower frequency cutoff for the components is all that
is required.

Note that the first (zero-frequency) component is effectively the mean
data value $\left<x(t)\right>$ (which can be pre-estimated and removed
in a robust way e.g.\ by taking a clipped median).  The presence of
gaps in the data provides us with a natural way of obtaining several
independent estimates of $\left<X_{\rm ls}(w)\right>$ by measuring it
separately in each interval between gaps, or alternatively provides a
natural boundary for doing independent light curve decompositions.

\begin{figure}
  \centering
  \epsfig{file=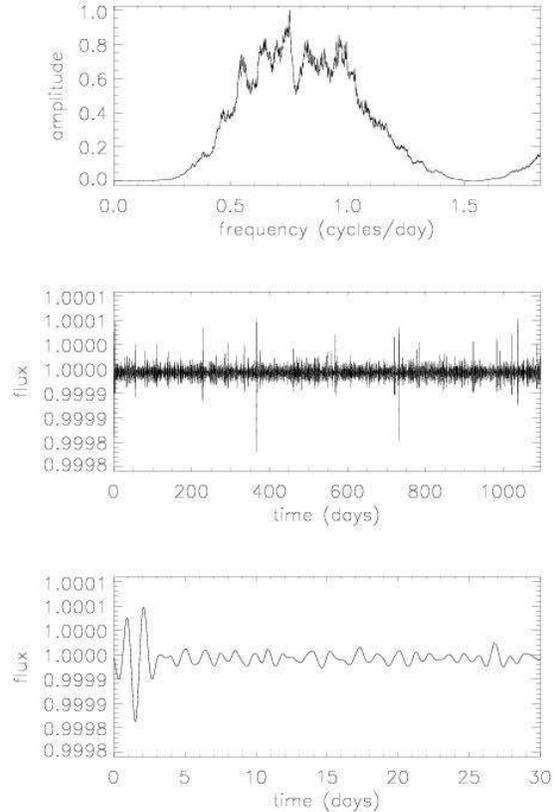, width = 0.9\linewidth}
  \caption{Top panel: Equivalent matched filter constructed using the
    light curve shown in Fig.~\ref{fig:lc} and a reference box-shaped
    transit of duration 0.65~d. Middle panel: Filtered light
    curve. Bottom panel: Filtered light curve, $1^{\rm st}$ 30 days,
    with a transit 1.5~d after the start.}
  \label{fig:lssd}
\end{figure}

Fig.~\ref{fig:lsfit} illustrates this least-squares fitting method, as
applied to the light curve shown in Fig.~\ref{fig:lc}. The top panel
shows ``power spectrum'', i.e.\ the coefficients $a_k$ versus
frequency, while the bottom two panels show the light curve
reconstructed by summing the fitted sine-curves. Note that high
frequency variations are not reconstructed as only the first 2000 sine
components were fitted (well below the Nyquist limit, but amply
sufficient for the purposes of following the long timescale stellar
variability).

The decomposition of the reference (transit) signal can usually be well
approximated analytically. For example if a simple box-shaped transit of
duration $d$ is adopted as reference signal, the $k^{\rm th}$
coefficient is given by:
\begin{equation}
  \label{eq:r_k}
  r_k = \frac{\sin \left( \pi k d / \delta t \right)}{\pi k d / \delta t}
\end{equation}
However, this decomposition can also be performed in the same way as
for the received data, for a reference signal of any given shape. The
sets of coefficients $a_k$ and $r_k$ then define the filter $h_k$,
which is equivalent to the Wiener, or matched filter of the previous section:
\begin{equation}
  \label{eq:h_k}
  h_k = \frac{\left<|r_k|^2\right>} {\left<|a_k|^2\right>} \hspace{1.5cm}
  h_k = \frac{\left<|r_k|\right>} {\left<|a_k|\right> }
\end{equation}
where the first expression corresponds to the standard Wiener filter,
and the second to the filter used in \citet{caf03}.

Fig.\ref{fig:lssd} illustrates this filtering method. Using the second
expression in Eq.~(\ref{eq:h_k}), a ``matched filter'' $h_k$ (top
panel) is constructed from the coefficients $a_k$ and $r_k$ (the
latter computed according to Eq.~\ref{eq:r_k}). The filtered light
curve, obtained by multiplying the $a_k$ by $h_k$ and reversing the
``transform'', is shown in the middle panel, with a zoom on the first
30 days in the bottom panel.

\begin{figure}
  \centering
  \epsfig{file=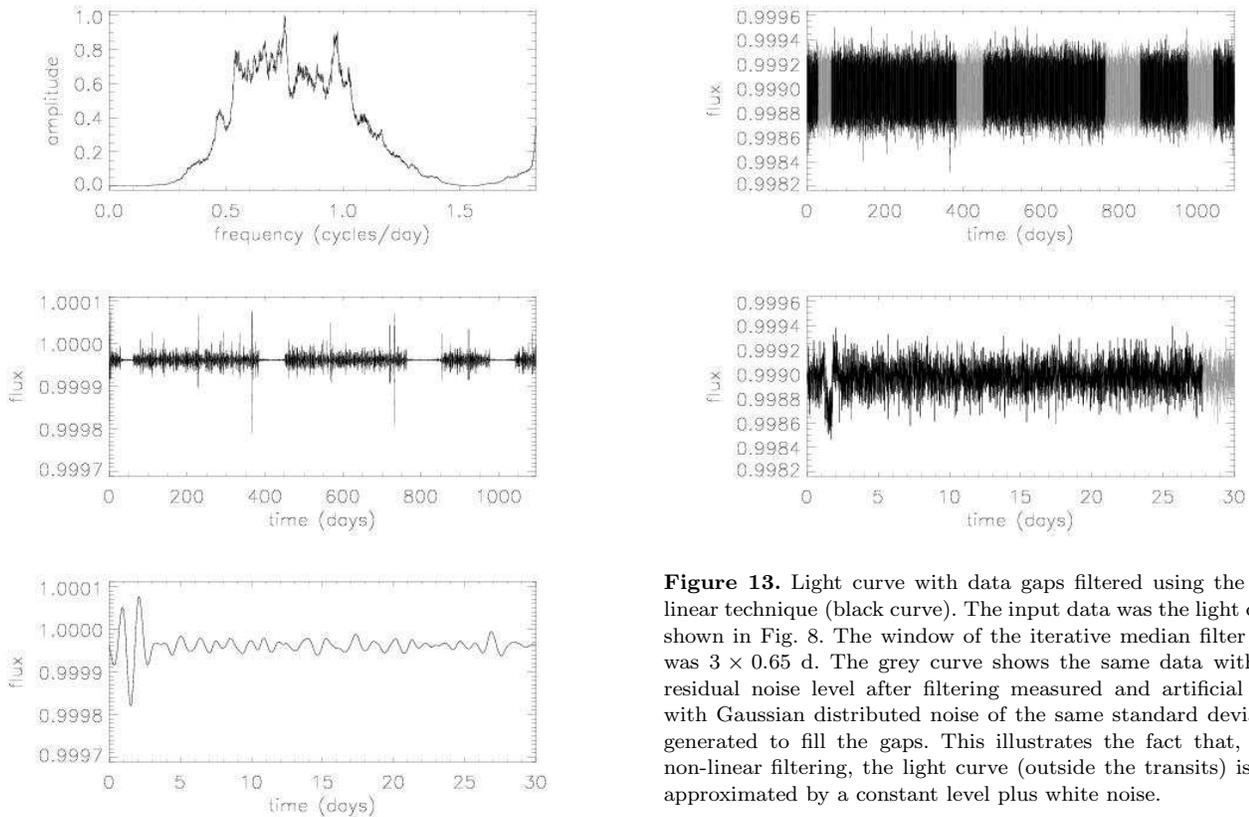, width = 0.9\linewidth}
  \caption{As Fig.~\ref{fig:lssd}, but the input light curve is that
    shown in Fig.~\ref{fig:lc_g}, with 4 significant data gaps.}
  \label{fig:lssd_g}
\end{figure}

Fig.~\ref{fig:lssd_g} shows the results of the matched filter
constructed using the least-squares fitting method when the light curve
contains gaps (as in Fig.~\ref{fig:lc_g}). The performance of the
filter is generally not affected by the gaps, though artifacts near
gap boundaries can sometimes be introduced.

The case of irregular sampling is not illustrated here, for practical
reasons: if the sampling was allowed to vary, say, by $\pm 10$\,\% of
the normal sampling time in a random fashion, the effect is not visible
in plots of such long light curves. In any case, we have found it to have
negligible effect on the the least-squares filtering.

\subsection{Non-linear filtering}
\label{sec:nlf}

If the timescale of the transits is shorter than the timescale for the
majority of the dominant stellar variations, iterative non-linear time
domain filters provide a powerful way of separating out short
timescale events. A good example of this type of approach can be
based around a standard median filter.

\begin{figure}
  \centering
  \epsfig{file=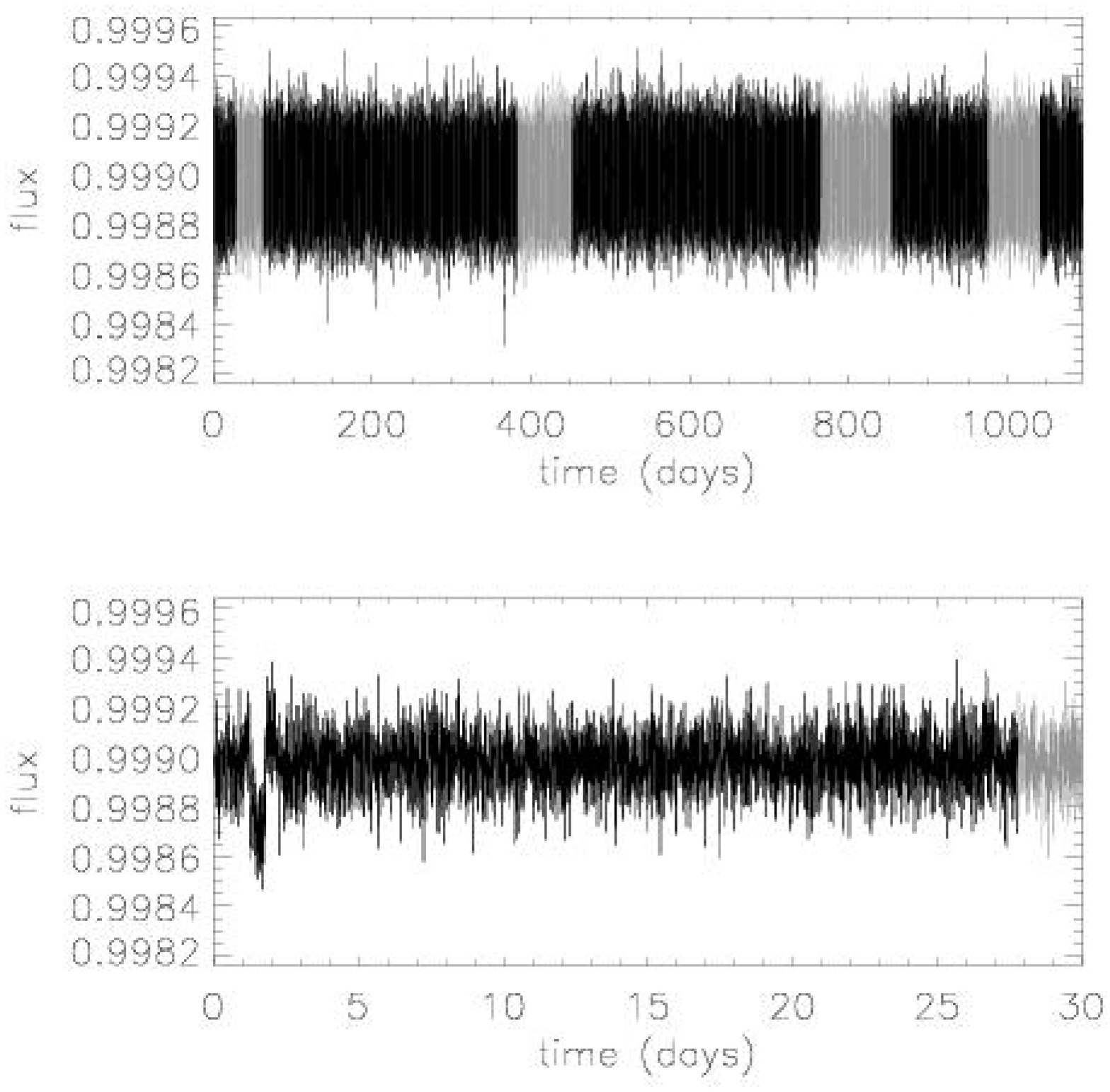, width=0.9\linewidth}
  \caption{Light curve with data gaps filtered using the non-linear
    technique (black curve). The input data was the light curve shown
    in Fig.~\ref{fig:lc_g}. The window of the iterative median filter
    used was $3 \times 0.65$~d. The grey curve shows the same data
    with the residual noise level after filtering measured and
    artificial data with Gaussian distributed noise of the same
    standard deviation generated to fill the gaps. This illustrates
    the fact that, after non-linear filtering, the light curve
    (outside the transits) is well approximated by a constant level
    plus white noise.}
  \label{fig:nl_g}
\end{figure}

The data is first, if necessary, split into segments, using any
significant gaps in temporal coverage to define the split
points. These gaps, defined as missing or bad data points, or
instances where two observations are separated in time by more than a
certain duration, can be automatically detected.

Each segment of data is then iteratively filtered using a median
filter of window $\sim 2$ to $3$ times the transit duration, followed
by a (small window) box-car filter to suppress level quantisation.
The difference between the filtered signal and the original is used to
compute the (robust) MAD-estimated scatter (sigma) of the
residuals. The original data segments are then $k$-sigma clipped (with
$k=3$) and the filtering repeated, with small gaps and subsequent
clipped values flagged and ignored during the median filtering
operation.  The procedure converges after only a few iterations.

Break points and/or edges are dealt with using the standard technique
of edge reflection to artificially construct temporary data
extensions.  This enables filtering to proceed out to the edges of all
the data windows.

The main advantage of using a non-linear filter is that the exact
shape of the transit is irrelevant and the only free parameter is the
typical scale size of the duration of the transit events.  The main
drawback is that the temporal information in the segments is
essentially ignored.  However, providing the sampling within segments
is not grossly irregular this has little impact in practice. This
filter is also relatively fast due to its simplicity: with the same
computer as before, the running time for a transit duration of $\sim
0.5$~day is 4~s per light curve, about the same as the time required
for the Wiener filter. The least-squares fitting method was
significantly slower (requiring approximately 30~s when 1500
frequencies were fitted).
 
Figure~\ref{fig:nl_g} illustrates this method as applied to the light
curve with gaps shown in Fig.~\ref{fig:lc_g}. As with the indirect
least-squares filtering, the high frequency noise remains, but this
does not impede transit detection. Given the simplicity of this method
and its good performance in the presence of data gaps, it appears to
be the most promising, as long as the sampling remains relatively
regular (if the sampling is significantly irregular, the least-squares
fitting method, which takes the time of each observation into account
directly, is likely to perform better).

The results of applying the transit search algorithm to the filtered
light curve are shown in Fig.~\ref{fig:trres}. The detection is
unambiguous (and remains so for a 1.5~$R_{\oplus}$ planet with
otherwise identical parameters, though the detection is not successful
for a 1~$R_{\odot}$ planet with only 3 transits\footnote{The star is a
4.5~Gyr old G2 dwarf in all cases.}.

\begin{figure}
  \centering
  \epsfig{file=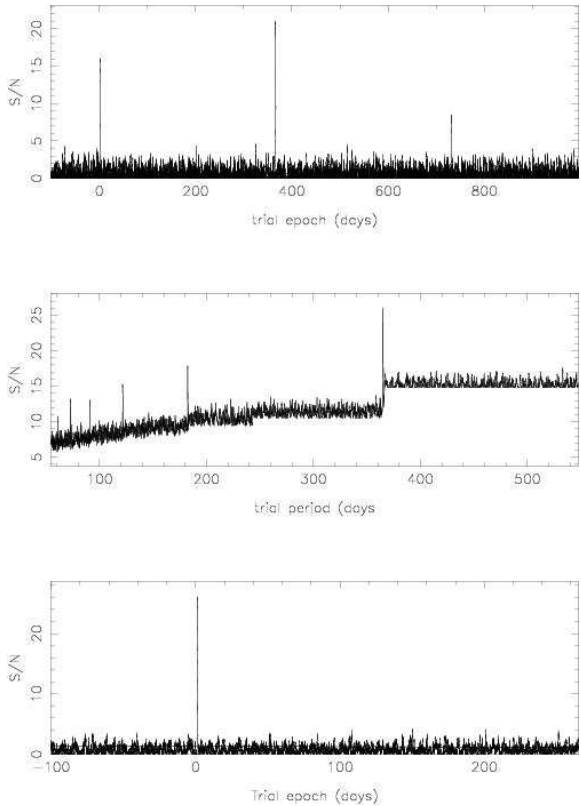, width=0.9\linewidth}
  \caption{Results of transit search after non-linear filtering. The
    input of the transit search program was the black curve shown in
    Fig.~\ref{fig:nl_g}. Top panel: detection statistic as a function
    of trial epoch for the preliminary single transit search (see
    Sect.~\ref{sec:speed}). The signature of all three transits ($e =
    1.5$, $366.5$ \& $731.5$~d) is clearly visible. Middle panel:
    multiple transit detection statistic as a function of trial
    period. Bottom panel: multiple transit detection statistic as a
    function of epoch at the optimal period of $365.0$~d. The detected
    epoch ($1.5$~d) is correct. The x-axis for the top and bottom
    panels were shifted by $100$~d for clarity.} 
  \label{fig:trres}
\end{figure}

The step-like appearance and systematic slope of the middle panel (period 
determination) is due to a combination of the discrete (and small) number of 
potential transits of the phase estimation stage which precedes it and the 
search for a minimum (over phase) at each trial period.  For each trial 
period the number of independent attempts to find a maximum in phase/epoch 
increases as the trial period increases.  Furthermore, the single transit 
phase has negative-going excursions clipped out to enhance the detectability
of real transits.  This leads to a systematic bias toward higher maxima as a 
function of trial period.  The steps are at harmonics and sub-harmonics of
the fundamental period and are due to quantisation of the number of possible
transits within each local trial period search.

The overall signal-to-noise ratio of the three combined
transits in the filtered light curves were approximately 26, 12
and 6 for planets of radius 2.0, 1.5 and 1.0~$R_{\oplus}$
respectively.  The fact that the 1.0~$R_{\oplus}$ case was not
detected is therefore roughly consistent with the SNR
limit of 6 stated by \citet{kzm02}.
 
\section{Performance evaluation}
\label{sec:perf}

In this section, we describe Monte Carlo simulations carried out
to evaluate the performance of the transit detection algorithm
described in Sect.~\ref{sec:chi2box}, combined with the iterative
non-linear filter introduced in Sect.~\ref{sec:nlf}.

\subsection{Method}
\label{sec:perfmet}

The method employed was identical to that described in Section 5.1 of
\citet{af02}, which was first used in the context of transit searches
by \citet{ddk+00}. The detection statistic (in this case the
signal-to-noise ratio of the best candidate transit) is computed for
$N$ light curves with transits. All light curves have the same
parameters, but different realisations of the noise and different
epochs randomly drawn from a uniform distribution (the epoch should
not affect the detection process). The process is repeated for $N$
transit-less light curves, which have noise characteristics identical
to those of the light curves with transits.  The chosen value of 100
for $N$ is a compromise between accuracy and time constraints, and
suffices to give a reasonable estimate of the performance of the method.

As the aim was to test the combined filtering and detection process,
the light curves were subjected to the iterative nonlinear filter,
before being forwarded to the transit detection algorithm. To avoid
prohibitively time-consuming simulations, and thus to allow several
star/planet configurations to be tested, a single transit duration
value was used (corresponding roughly to the FWHM of the input
transits).

Once the algorithm has been run on all the light curves, the next step
consists in choosing a detection threshold: any light curve for which
the maximum detection statistic exceeds this threshold will be
considered to contain a candidate transit. If a transit-less light
curve gives rise to a statistic above the threshold, a false positive: 
a candidate transit appears to have been detected when there
is in fact none. Conversely, if the maximum detection statistic for a
light curve with transits lies below the threshold, the transit(s) will
go undetected: a false negative.

The optimal threshold, given a set of light curves which are known to
share the same noise characteristics, can be chosen from the results
of the transit search itself to minimise the false alarms and missed
transits. This is illustrated in a schematic way in Figure~3 of
\citet{af02}.   Detection statistic histograms ideally should show a 
clear separation between real transits and false alarms, allowing a
simple choice of boundary between the respective distributions.
The location of the boundary is chosen as a compromise between
maximising the detection rate and minimising the number of false alarms.

In certain circumstances, it might be more important to minimise
missed detections (for example if the sought-after events are very
rare, particularly if false alarms can easily be weeded out at a later
stage). In other circumstances (for example if it is very difficult to
test the reliability of any candidate events through further
observations) it may be more desirable to minimise false alarms.
However, as our present aim is simply to carry out a simple
performance evaluation, we did not give priority to either kind of
error over the other and just minimised the sum of the two types of
error.

\subsection{Results}
\label{sec:perfres}

\begin{figure}
  \centering 
  \epsfig{file=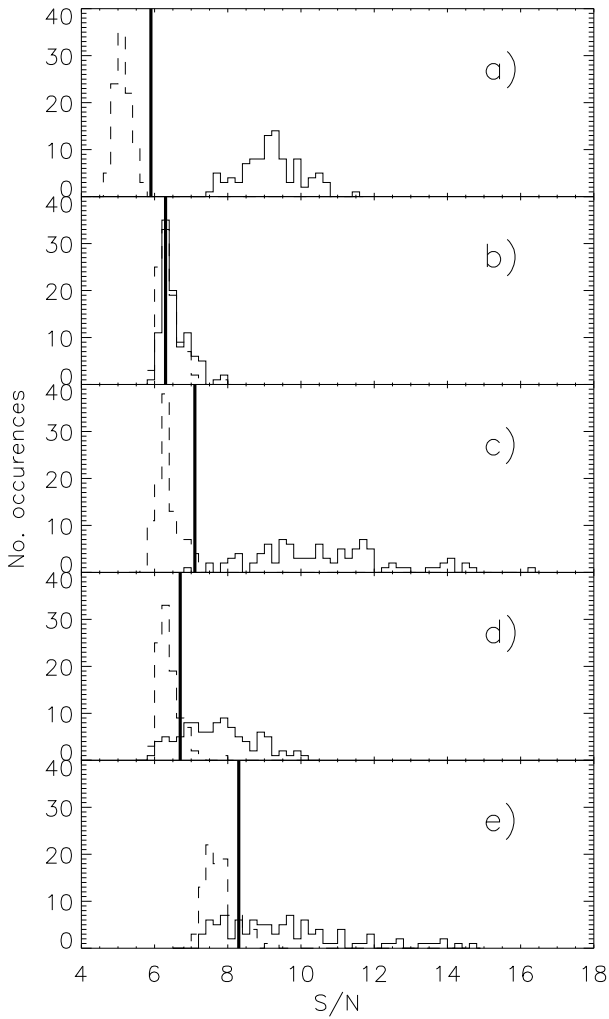, width=8cm, bbllx=58, bblly=36, bburx=328, bbury=498, clip=}
  \caption{Results of the performance evaluation for 5 star/planet
    configurations, as detailed in Table~\ref{tab:hdist}. Solid
    histograms: distributions of the maxima of S/N statistics computed
    by the transit detection algorithm after non-linear filtering for
    100 light curves containing transits. Dashed histograms: idem for
    100 light curves containing stellar variability and photon noise
    only. Thick vertical lines: optimal detection threshold.}
  \label{fig:hdist}
\end{figure}

\subsubsection{Photon-noise only case}
\label{sec:photonly}

The aim of this simulation was to compare the performance of the
present algorithm to others, which have mostly been tested on
white-noise only light curves.

If the present method is to improve on the performance of the Bayesian
approach it is derived from, it should be able to detect reliably a
$1.0~R_{\oplus}$ planet orbiting a G2V star with photon noise
corresponding\footnote{for \emph{Eddington}'s expected photometric
performance} to $V=13$, given three transits in the light curve.  In
\citet{af02}, simulations showed that such a planet should be easily
detected around a smaller (K5V) but fainter ($V=14$) star with the
older algorithm and no filtering. The $V=13$, G2 case corresponds to a
$S/N$ that is larger by a factor of 1.07, and should therefore be
detected easily if the new method is as efficient as the old.

\begin{table}[h]
  \centering
  \caption{Light curve characteristics for each panel of Fig.~\ref{fig:hdist}.}
  \label{tab:hdist}
  \begin{tabular}{lccccc}
    \footnotesize
    Panel & a) & b) & c) & d) & e) \\
    \hline
    Photon noise & $\surd$ & $\surd$ & $\surd$ & $\surd$ & $\surd$ \\
    Stellar & $\times$ & $\surd$ &  $\surd$ &  $\surd$ &  $\surd$ \\
    variability &&&&&\\
    Age (Gyr) & 4.5 & 4.5 & 4.5 & 4.5 & 4.5 \\
    SpT & G2V & G2V & G2V & G2V & G2V \\
    $R_{\rm pl}$ ($R_{\oplus}$) & 1.0 & 1.0 & 1.5 & 1.0 & 1.0 \\
    Period (yr) & 1.0 & 1.0 & 1.0 & 0.5 & 1.0
  \end{tabular}
\end{table}

After a set of simulations was run for such a configuration, the
maximum detection statistic from the noise only light curves was
$S/N=5.79$, while the minimum value from the light curves with
transits was $S/N=7.41$ (see Fig.~\ref{fig:hdist}a). Any threshold in
between would therefore allow the detection of all the transits where
present, with no false alarms.

Note that the $S/N$ limit of 6, quoted by \citet{kzm02} for their BLS
method, which is statistically close to ours, falls as expected in the
range of thresholds that would be suitable in the present case.

\subsubsection{Photon noise and stellar variability}
\label{sec:borderline}

$\bullet$ {\bf $1.0~R_{\oplus}$ planet orbiting a G2V star}\\ 
\noindent This configuration is identical to that in
Sect.~\ref{sec:photonly}, but with stellar variability added. It is
also similar to the case illustrated in Figs.~\ref{fig:lc} to
\ref{fig:trres}, but with a smaller planet. The results are shown in
Fig.~\ref{fig:hdist}b.  The distributions of the detection statistics
from the light curves with and without transits overlap almost
entirely, i.e.\ the performance is poor. The threshold that minimises
the sum of false alarms and missed detection leads to 56 of the first
and 26 of the second.

Assuming that the sampling rate, light curve duration, and stellar
apparent magnitude are fixed, there are three factors which should
lead to better performance: a larger planet, a shorter orbital period
(i.e.\ more transits) or a smaller star. Each of these options in turn
is investigated below.

\noindent $\bullet$ {\bf $1.5~R_{\oplus}$ planet orbiting a G2V star}\\
\noindent The histograms are relatively well separated (see
Fig.~\ref{fig:hdist}c), with only a small overlap, so that the optimal
threshold of $S/N=7.85$ lead to one missed detection and no false
alarms.

It is interesting to note the similarity between the results of this
simulation and the requirements used for the design of the
\emph{Kepler} mission, which was to detect planets given a
signal-to-noise ratio totalling at least 8 for at least three
transits\footnote{See {\tt www.kepler.arc.nasa.gov/sizes.html}.}.

\noindent $\bullet$ {\bf $1.0~R_{\oplus}$ planet orbiting a G2V star
with 6 transits}\\
\noindent The aim of this set of simulations was to investigate the
effect of increasing the number of transits in the light curve by a
factor of two by reducing the orbital period to 182~d. This is
equivalent to increasing the overall duration of the observations. As
expected, this leads to higher $S/N$ values and hence better
performance, with only 13 false alarms and 16 missed detections (see
Fig.~\ref{fig:hdist}d).

\noindent $\bullet$ {\bf $1.0~R_{\oplus}$ planet orbiting a K5V star}\\
\noindent A K5 star is smaller than a G2 star, leading to deeper
transits, but also more active, leading to more stellar
variability. Recent studies\citep{afg04} suggested that the former
effect prevailed over the latter, and that K or even M type stars
might make better targets for space missions seeking to detect
habitable planets than G stars, but these were based only on results
from a few individual light curves, rather than Monte Carlo
simulations.

The present tests confirm this trend: the separation between the with-
and without transit distributions is wider (see Fig.~\ref{fig:hdist}e)
than in the previous case, though the best-threshold false alarm and
missed detection rates remain high at 13 and 25\,\% respectively.

Note the higher $S/N$ values for the transit-less light curves
compared to the G2 case, which suggests the presence of more residual
stellar variability after filtering, as would be expected.

\section{Discussion}
\label{sec:discus}

Starting from a general purpose maximum likelihood approach we have
demonstrated the the links between a variety of period and transit 
finding methods and have shown that matched filters, cross-correlation, 
least-squares fitting and maximum likelihood methods are all facets of 
the same underlying principle.  In the simple approximation of 
rectangular-shaped transits embedded on a flat continuum and in white 
noise, all of these approaches can be tuned to give similar detection 
results.

The transit detection algorithm presented here provides a unified
approach linking all these methods. Computational efficiency is of
particular importance in the context of large, long duration, high
sampling missions such as \emph{Eddington} and \emph{Kepler}, and the
present method would allow a search for transits by habitable planets
to be performed on 20,000 3~yr long light curves with 10~min sampling
in less than a day. Including the time required to apply the nonlinear
filter, which for the laptop used takes $\sim 4\,$s per light curve
per filter duration, this would increase to $\sim 3\,$d (using three
different filter durations). This is achieved at no cost in
efficiency: in white noise only, the algorithm is capable of detecting
transits down to approximately the same $S/N$ limit as that quoted by
\citet{kzm02} for their BLS method, which has been the most successful
method to date in terms of practical results, being used by the OGLE
team to discover most of their candidate transits, see
\citep{uzs+02,ups+03}.

This approach is predicated on the assumption of periodic transits
hidden in random noise, usually assumed to be superposed on a flat
continuum with regular continuous sampling.  In the real world,
stellar (micro) variability is expected to be the dominant signal
component.  We have then shown how to generalise the transit finding
method to the more realistic scenario where complex stellar
variability, irregular sampling and long gaps in the data, are all
present.

The two filtering methods developed to deal with this case share some
advantages -- both can be applied to data with gaps -- but they also
have different properties.  The least-squares fitting method is
capable of making use of the time information in data with irregular
sampling. It also allows a theoretically optimal filter (i.e.\ the
Wiener or matched filter) to be combined with a pre-whitening filter,
although from the point of view of detection, the matched filter
is the main active component of any maximum likelihood-based
detection algorithm.  As a by product of the filtering, the stellar
signal can also be reconstructed. However, this is computationally
intensive, particularly if one wishes to fit higher frequencies. Its
performance also depends quite critically on concordance between the
duration of the reference transit and that of any true transit.

On the other hand, iterative non-linear filtering is simple to
implement and fast, but ignores any local time information (except for
the long gaps which are detected automatically). This means that its
performance is likely to degrade if the sampling is seriously
irregular.  However, it is the most efficient method in cases such as
those investigated here.  By removing any signal on timescales longer
than two-three times the estimated transit duration, it is likely to
be less affected by the value chosen for that duration.  Although more
work is needed to establish quantitatively the relative merits of the
two approaches, it seems more efficient, given the results so far, to
use the iterative non-linear filtering method prior to a general
transit search. The least-squares fitting method could be employed in
the more difficult (e.g.\ very irregular sampling) or borderline (as
in Sect.~\ref{sec:borderline}) cases, where the additional information
used about the transit shape may lead to better performance.

Whatever the method used, there is a fundamental limit to what can be
achieved. Stellar variability can only be filtered out if an
orthogonal decomposition of the transit and stellar signal is
possible, e.g.\ if the two signatures in the frequency domain do not
overlap by too much. Therefore, very rapidly rotating stars where the
rotation period is close to the transit duration, or stars showing
much more power than the Sun on timescales of minutes to hours (e.g.\
higher meso- or super-granulation) will be problematic targets. Even
in the hypothetical situation where all stellar noise is removed, the
remaining white noise will also place a limit on the performance of
the transit detection algorithm, and hence on the apparent magnitude
of star around which transits of a certain depth can be found.  In
white Gaussian noise, any transit yielding a signal-to-noise ratio
above a fixed threshold (estimated to be $\approx 6$ in
Sect.~\ref{sec:photonly}) should be detectable. Considering photon
noise alone, for a given stellar radius, orbital period and transit
duration, the smallest detectable planet radius would therefore scale
as $B^{-1/4}$ or $\exp(m/10)$ where $B$ and $m$ are the star's
apparent brightness and magnitude respectively.

The natural progression of this work will be further
quantification of the performances attained, and the identification of
the best method to use for a given situation (i.e.\ star-planet
combination, instrument characteristics and/or sampling). As
in the present paper, this can be done through Monte Carlo
simulations, and more realistic noise profiles can be included in the
light curves (e.g.\ instrumental noise). Extensive simulations can be
performed for a given target field by coupling the stellar variability
model to a galactic population model and any available extinction
information on the field. However, it will only be meaningful to carry
out such simulations when the design, target fields and observing
strategies of the missions in question are finalised and when more
information about stellar micro-variability is available.

Our main conclusion if that even with realistic contamination from
stellar variability, irregular sampling, and gaps in the data record,
it is still possible to detect transiting planets with an efficiency
close to the idealised theoretical bound. In particular, space missions
are tantalisingly close to being capable of detecting earth-like planets
around G and K dwarfs.

\section*{Acknowledgements}

S.\ A. acknowledges support from PPARC studentship number
PPA/S/S/2003/03183 and from the Isaac Newton Trust. We are grateful to
F.\ Favata and G.\ Gilmore for their careful reading of the manuscript
and helpful comments.

\bibliographystyle{mn2e} \bibliography{MD1306rv}

\appendix

\section[]{Simulation of realistic \emph{Eddington} light curves}
\label{sec:ex}

In this appendix we briefly outline the method used to simulate the
light curves shown in Figs.~\ref{fig:lc} \& \ref{fig:lc_g}.

\subsection{Planetary transits}
\label{sec:pl}

\citet{dee99}'s IDL based Universal Transit Modeller (UTM) was used to
simulate noise-free light curves. UTM includes a linear limb-darkening
law, and limb-darkening coefficients from \citet{ham93} were used. For
a given star-planet configuration, the other input parameters were the
ratio of planetary to stellar radius, the planet's orbital period and
distance, and the sampling time and duration. For the latter, values
of 10~min and 3~yr respectively were used, as appropriate for
\emph{Eddington} in planet-finding mode \citep{fav03}. The output is
in units of relative flux, normalised to an out-of-transit value of
1.0. These units are used throughout. Note that no reflected light
from the planet is included, and that all orbits are assumed to be circular. 
The planet's orbital plane is also assumed to be aligned along the 
line-of-sight.

For the current paper, we chose to model a $2~R_{\oplus}$ planet
orbiting a G2V star ($R_{\star}=1.03~R_{\odot}$), i.e.\ a radius ratio
of $0.018$, leading to a relative transit depth of $3.24 \times
10^{-4}$. This is not the smallest detectable planet around such a
star (with the methods presented here), but it is the smallest for
which transits are visible by eye in both the pre- and post-filtering
light curves. The planet's orbital period is 1~yr, and its orbital
distance 1~AU. The epoch of the first transit is 1.5~d. The power
spectrum of this transit-only light curve is shown as the black line
with repeated ``humps'' in Fig.~\ref{fig:psnspec}.

\subsection{Intrinsic stellar variability}
\label{sec:st}

The model used to simulate stellar micro-variability, which allows the
generation of light curves for stars of various spectral types and
ages, was presented in detail in \citet{afg04}, with the aim of
testing and refining filtering and transit detection algorithms, in
the context of space-based transit searches such as COROT,
\emph{Eddington} and \emph{Kepler}.

The starting point for the model is the Sun's photometric variability,
which has been studied at ultra-high precision since January 1996 by
the VIRGO experiment \citep{faa+97} onboard the SoHO
observatory. Empirical scaling laws, either published
\citep{sku72,nhb+84} or derived from published datasets
(\citealt{rtl+87,rls+95,rls+98,hbd+00}, for a wide range of stars),
are then used to scale the amplitude and frequency distribution of the
Sun's variability to other stellar ages and masses.

Light curves can be generated for dwarfs of any spectral type between
F5 and K5, and for all ages later than the Hyades (625 Myr,
\citealt{pbl+98}). In the present paper, a 4.5~Gyr old G2V star was
modelled, again with a sampling time of 10~min and duration of
3~yr. The stellar light curve, also in relative flux units (and whose
power spectrum is shown as the lower grey line in Fig.~\ref{fig:psnspec}),
is then multiplied by the planetary light curve described in \ref{sec:pl}.

The IDL source code used to construct these, together with a number of
existing simulated light curves, are available from 
\verb\www.ast.cam.ac.uk/~suz/simlc\.

\subsection{Photon noise}

The \emph{Eddington} baseline configuration\footnote{\small \tt
astro.estec.esa.nl/Eddington/Tempo/eddiconfig.html}, at the time of
writing, consists of four co-aligned wide-field telescopes, with a
total collecting area of $0.764~{\rm m}^2$.  Combined with the optics
and CCD performance, this leads to an expected photon count of, for
example, $1.4 \times 10^5~\gamma~{\rm s}^{-1}$ for a $V=13$ star. The
photon noise in relative flux units should thus be well approximated
by a Gaussian distribution with a normalised standard deviation of $1.09 \times
10^{-4}$ for 10~min integrations, and such a randomly generated photon
noise value was added to each data point in the combined star-planet
light curve. The result is the light curve shown in Fig.~\ref{fig:lc},
while the power spectrum of the noise component is shown as the
approximately constant black line in Fig.~\ref{fig:psnspec}.

\subsection{The above with gaps}

To investigate the impact of data gaps, the following four sections of
data were removed from the gap-less light curve:
\begin{itemize}
\item indices 4\,000 to 8\,999 (i.e.\ $t=27.8$ to 62.5~d);
\item indices 55\,092 to 65\,060 (i.e.\ $t=382.6$ to 451.8~d);
\item indices 110\,000 to 123\,009 (i.e.\ $t=763.9$ to 854.2~d);
\item indices 140\,395 to 149\,999 (i.e.\ $t=975.0$ to 1041.7~d).
\end{itemize}

These were chosen arbitrarily, but with the aim of ensuring a variety
of gap and data interval durations, and avoiding the removal of any
transits. In reality, data gaps are of course likely to affect the
number of observed transits, but this is a different issue to that
investigated here, i.e.\ the development of filters which can remove
the stellar signal in the presence of gaps regardless of the presence
(or lack) of transits.  The resulting light curve is shown in
Fig.~\ref{fig:lc_g}.  

Note that missions like \emph{Eddington} are expected to have a very
high duty cycle ($>95$\,\% -- Favata, priv.\ comm.), compared to a
value of $\sim 70$\,\% for the simulated light curve used in the
present work. Such a low duty cycle is therefore even more
conservative than the expected worst case scenario.

\bsp

\label{lastpage}

\end{document}